\documentclass[conference]{IEEEtran}
\IEEEoverridecommandlockouts
\usepackage{cite}
\usepackage{amsmath,amssymb,amsfonts}
\usepackage{algorithmic}
\usepackage{graphicx}
\usepackage{textcomp}
\usepackage{tabularx}
\usepackage{booktabs}
\usepackage{multirow}
\usepackage[compatibility=false]{caption} 
\usepackage{graphicx} 
\usepackage{amsmath} 
\usepackage{amsfonts} 
\usepackage{url} 
\usepackage{siunitx} 
\usepackage[table]{xcolor} 
\usepackage{adjustbox} 
\def\BibTeX{{\rm B\kern-.05em{\sc i\kern-.025em b}\kern-.08em
    T\kern-.1667em\lower.7ex\hbox{E}\kern-.125emX}}
\begin{document}

\title{Poster: Enhancing GNN Robustness for Network Intrusion Detection via Agent-based Analysis \\
}

\author{\IEEEauthorblockN{ Zhonghao Zhan}
\IEEEauthorblockA{\textit{Department of Computing} \\
\textit{Imperial College London}\\
}
\and
\IEEEauthorblockN{Huichi Zhou}
\IEEEauthorblockA{\textit{Department of Computing} \\
\textit{Imperial College London}\\
}
\and
\IEEEauthorblockN{Hamed Haddadi}
\IEEEauthorblockA{\textit{Department of Computing} \\
\textit{Imperial College London}\\
}

}

\maketitle

\begin{abstract}
Graph Neural Networks (GNNs) show great promise for Network Intrusion Detection Systems (NIDS), particularly in IoT environments, but suffer performance degradation due to distribution drift and lack robustness against realistic adversarial attacks. Current robustness evaluations often rely on unrealistic synthetic perturbations and lack demonstrations on systematic analysis of different kinds of adversarial attack, which encompass both black-box and white-box scenarios. This work proposes a novel approach to enhance GNN robustness and generalization by employing Large Language Models (LLMs) in an agentic pipeline as simulated cybersecurity expert agents. These agents scrutinize graph structures derived from network flow data, identifying and potentially mitigating suspicious or adversarially perturbed elements before GNN processing. Our experiments, using a framework designed for realistic evaluation and testing with a variety of adversarial attacks including a dataset collected from physical testbed experiments, demonstrate that integrating LLM analysis can significantly improve the resilience of GNN-based NIDS against challenges, showcasing the potential of LLM agent as a complementary layer in intrusion detection architectures.
\end{abstract}


\section{Introduction}
Graph Neural Networks (GNNs) offer significant potential for Network Intrusion Detection Systems (NIDS) by modeling complex network relationships. However, realizing their practical utility requires addressing critical robustness challenges often overlooked in standard evaluations. Our work tackles two primary issues revealed through rigorous preliminary analysis.

First, we address the challenge of generalization under distribution drift. Standard GNN evaluation often relies on training and testing within single, static datasets (e.g., a specific version of CIC-IDS or UNSW-NB15). This does not capture performance in real-world networks where traffic patterns constantly evolve \cite{andresini2021insomnia}. To investigate this, we constructed a \textit{unified dataset} by merging and standardizing flows from various canonical sources, including UNSW-NB15 \cite{moustafa2015unsw}, CIC-IDS2018 \cite{sharafaldin2018toward}, and Bot-IoT \cite{koroniotis2019towards}. Evaluating representative GNN models (such as EGraphSage \cite{lo2022graphsage}, Anomal-E \cite{caville2022anomal}, and CAGN-GAT \cite{jahin2025cagn}) using this unified dataset revealed a notable performance degradation compared to their single-dataset benchmarks, empirically confirming their susceptibility to distribution drift, a fundamental robustness problem.

Second, beyond drift, we examine robustness against adversarial attacks under realistic conditions. While GNNs might claim robustness, evaluations often use synthetic attacks, particularly white-box methods, that grant attackers unrealistic capabilities \cite{ma2020towards}. To probe vulnerability more realistically, we simulated practical, problem-space black-box attacks (such as node injection) on our unified dataset. These experiments demonstrated \textit{further} significant performance degradation in the evaluated GNN models, exposing their vulnerability even to attacks feasible for real adversaries. This finding underscores the inadequacy of current robustness claims and highlights the need for both more realistic adversarial testing scenarios and effective mitigation strategies.

The observed GNN fragility against both drift and realistic attacks motivates our core proposal: leveraging LLMs as simulated cybersecurity experts. Given that human experts can often identify sophisticated attacks or anomalies missed by automated systems \cite{ding2021iotsafe, siboni2016security}, we explore using LLM agents to perform preemptive analysis on network flow graph data. By scrutinizing graph structures and flow patterns, the LLM agent aims to filter threats or alert operators before they compromise the GNN. This poster presents our investigation into this LLM-agent approach, detailing its integration and presenting experimental results that demonstrate its effectiveness in enhancing GNN resilience against simulated realistic attacks, offering a promising path towards more robust NIDS deployment.

\begin{figure*}[t!]
  \centering
  \includegraphics[width=0.75\linewidth]{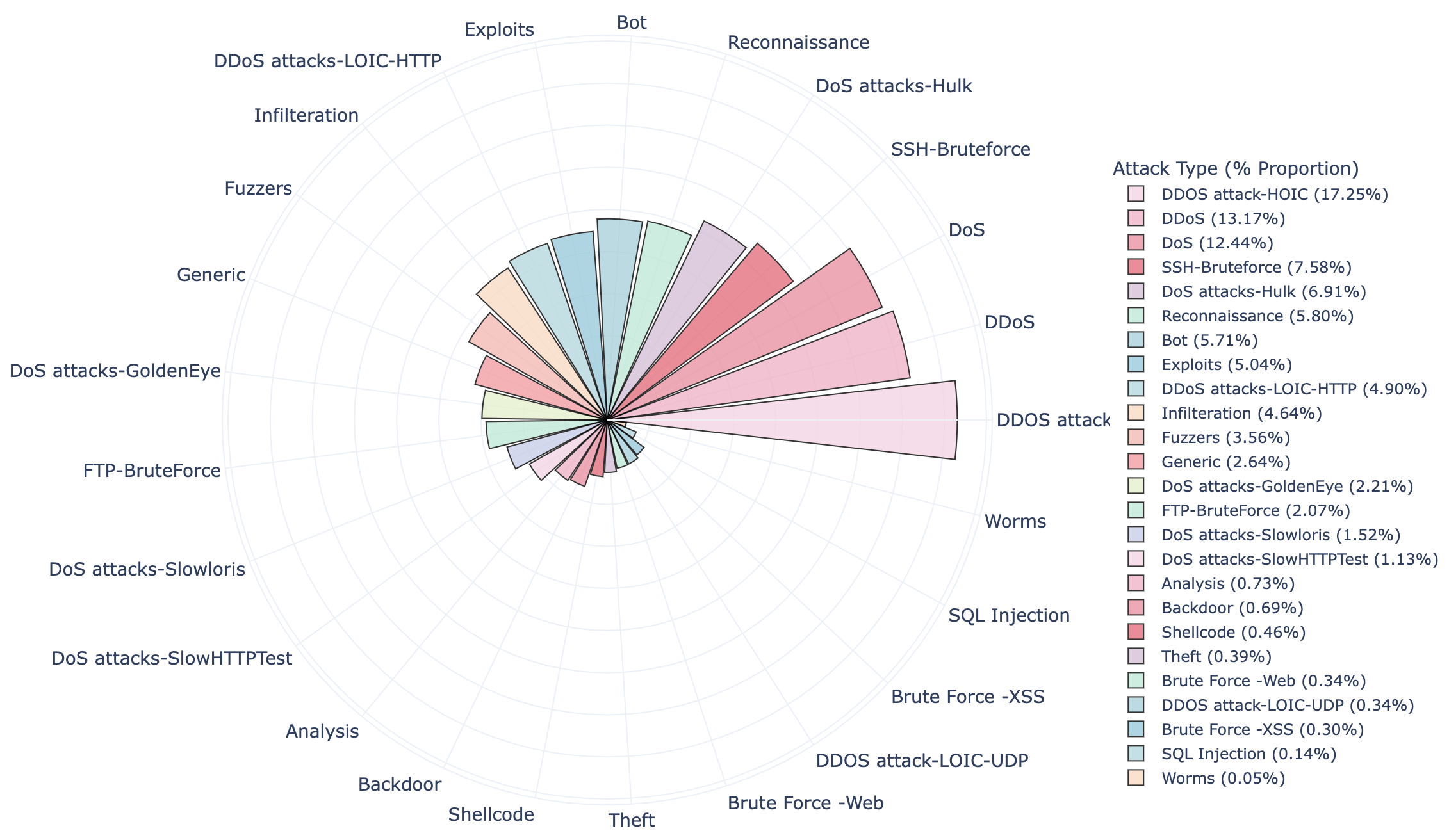}
  \label{fig: overview}
  \caption{The intrusion types of unified dataset for distribution drift test}
\end{figure*}

\section{Related Work}
Current evaluations of GNN-based NIDS face limitations hindering realistic assessment. Models are often evaluated on single static datasets, ignoring performance degradation due to distribution drift \cite{andresini2021insomnia}. Furthermore, adversarial robustness assessments frequently rely on synthetic perturbations generated without considering network domain constraints \cite{ma2020towards}, potentially rendering attacks impractical \cite{ma2020towards}. Much GNN robustness research uses generic graph benchmarks rather than NIDS-specific structures \cite{wang2025we}, and realistic structural attacks remain underexplored \cite{ma2020towards}. Existing GNN robustness benchmarks may also lack sufficient integration of drift or domain-specific realism \cite{zheng2021graph}. Ideas leveraging LLMs for robust evaluation or enhancing GNNs are emerging \cite{zhou2025trustrag, zhang2024can}, but applying them as expert simulators for NIDS remains unexplored. This evaluation gap leads to an overestimation of GNN resilience.

\begin{table}[htbp]
\centering
\caption{Comparison of GNN Performance Across Datasets}
\label{tab:gnn_comparison}
\small
\renewcommand{\arraystretch}{1.2}
\setlength{\tabcolsep}{4pt}
\sisetup{
  detect-weight        = true,
  detect-family        = true,
  round-mode           = places,
  round-precision      = 3,
  table-number-alignment = center
}
\begin{adjustbox}{max width=\columnwidth}
\begin{tabular}{
  @{} l l 
  S[table-format=1.3]    
  S[table-format=1.3]    
  @{}
}
\toprule
\textbf{Model} & \textbf{Dataset} & {\textbf{Accuracy}} & {\textbf{F1}} \\
\midrule
CAGN & Unified & 0.851 & 0.823 \\
     & UNSW-NB15 & 0.995* & 0.918* \\
     & CICIDS2017 & 0.975* & 0.881* \\
\midrule
AnomalE & Unified & 0.861 & 0.875 \\
        & UNSW-NB15 & 0.987* & 0.924* \\
        & CICIDS2018 & 0.971* & 0.924* \\
\midrule
E-GraphSage & Unified & 0.675 & 0.793 \\
            & NF-ToN-IoT & 0.672* & 0.810* \\ 
            & NF-Bot-IoT & 0.782* & 0.630* \\
\bottomrule
\multicolumn{4}{@{}p{\linewidth}@{}}{* Results reported in the original paper of the models in the table. The 'Unified' results reflect performance on the unified dataset used in this study. AnomalE F1 for Combined are weighted averages.} \\
\end{tabular}
\end{adjustbox}
\end{table}
\section{Methodology}
Our approach integrates LLM analysis into the GNN-based NIDS pipeline.

\subsection{Dataset and GNN Models}
We utilized a unified dataset constructed by merging and standardizing multiple NetFlow-based intrusion detection datasets (NF-BoT-IoT \cite{koroniotis2019towards}, NF-CSE-CIC-IDS2018 \cite{sharafaldin2018toward}, NF-UNSW-NB15 \cite{moustafa2015unsw}) to simulate distribution drift. This allows for evaluating GNN generalization across diverse network scenarios. We evaluated several GNN architectures commonly used in NIDS, including E-GraphSAGE \cite{lo2022graphsage}, Anomal-E \cite{caville2022anomal}, and CAGN-GAT \cite{jahin2025cagn}. Network flows were transformed into IP-centric communication graphs where nodes represent IPs and edges represent flows with associated features.

\subsection{LLM Agent Integration}
We designed an LLM-based mitigation strategy where LLM agents analyze elements of the network graph before GNN processing. This involves:
\begin{itemize}
    \item \textbf{Task Definition:} The LLM's task is to assess the relevance or potential maliciousness of graph components, such as injected nodes simulating certain attacks or suspicious edges/flows (conceptualized in Figure 2).
    \item \textbf{Prompting Strategy:} LLMs are prompted with textual information representing graph elements (e.g., descriptions of node interactions or flow characteristics) and asked to provide an analysis and relevance score.
    \item \textbf{Workflow:} The LLM acts as a filter or expert advisor. Based on the LLM's assessment, suspicious graph elements could be flagged, removed, or weighted differently before being fed into the GNN model. We specifically tested this against node injection attacks where the LLM attempts to identify the injected malicious nodes.
\end{itemize}

\subsection{Evaluation Protocol}
We evaluated the GNN models under several conditions:
\begin{enumerate}
    \item Baseline performance on standardized datasets.
    \item Performance under simulated distribution drift using the unified dataset.
    \item Robustness against synthetic attacks (PGD feature attacks, Edge Removal, Node Injection).
    \item Performance of the GNN with LLM-based mitigation applied, particularly focusing on the recovery from node injection attacks.
\end{enumerate}
Metrics included Accuracy, F1-Score, Precision, Recall, and AUC, with a focus on F1-Score due to class imbalance inherent in NIDS datasets.

\begin{figure}[t!] 
  \centering
  \includegraphics[width=0.9\linewidth]{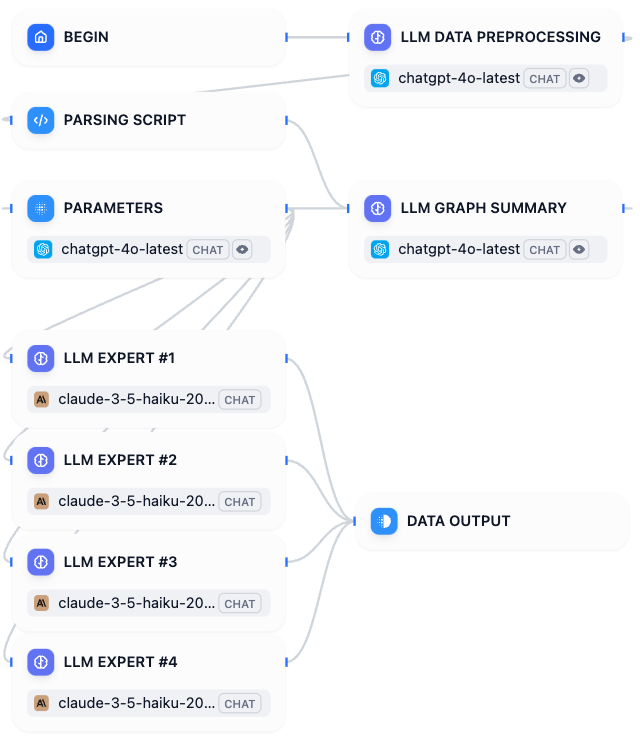} 
  \caption{Design diagram of the LLM Mitigation pipeline}
  \label{fig:Pi} 
\end{figure}

\section{Results and Discussion}
Our experiments confirm the susceptibility of standard GNN models to distribution drift and synthetic attacks. Node injection, in particular, is observed to degrade accuracy and F1-scores across the evaluated models.

The key finding highlighted in this poster is the effectiveness of the LLM mitigation strategy. LLMs (specifically tested models like GPT-4o and LLaMA 4) demonstrated a strong capability to identify artificially injected malicious nodes within the graph structure derived from netflow data, as shown in Table \ref{tab:llm_mitigation_results}. For instance, in scenarios with 20\% node injection (200 injected nodes into a base graph of 1000), leading LLMs correctly flagged a high percentage of these malicious nodes.

By filtering or identifying these malicious nodes based on LLM analysis, the downstream GNN classification performance improved significantly compared to the attacked scenario without LLM mitigation (Table \ref{tab:llm_mitigation_results}). While tested against synthetic node injection, the LLM's ability to reason about network interactions based on provided data hints at potential benefits for handling novel, zero-day attacks or complex drift patterns that manifest as unusual graph structures.

\begin{table}[htbp] 
\centering
\caption{Performance of LLMs in graph mitigation. CAGN-GAT Fusion\cite{jahin2025cagn} 
 is being tested. CF = Correctly Flagged, IF = Incorrectly Flagged}
\label{tab:llm_mitigation_results} 
\small 
\renewcommand{\arraystretch}{1.2}
\setlength{\tabcolsep}{4pt}
\sisetup{
  detect-weight        = true,
  detect-family        = true,
  round-mode           = places,
  round-precision      = 3,
  table-number-alignment = center
}
\begin{adjustbox}{max width=\columnwidth} 
\begin{tabular}{
  @{} l
  S[table-format=1.3]    
  S[table-format=1.3]    
  S[table-format=1.3]    
  S[round-precision=0, table-format=4.0]  
  S[round-precision=0, table-format=3.0]  
  S[round-precision=0, table-format=3.0]  
  @{}
}
\toprule
\textbf{Model}
  & {\textbf{Accuracy}}
  & {\textbf{F1}}
  & {\textbf{LLM Recall}}
  & {\textbf{Nodes}}
  & {\textbf{CF}}
  & {\textbf{IF}} \\
\midrule
Clean
  & 0.842 & 0.821 & {--} & 1000 & {--} & {--} \\
Claude 3.5 Haiku
  & 0.777 & 0.673 & 0.698 & 1036 & 0 & 164 \\
Claude 3.7 Sonnet
  & 0.774 & 0.728 & 0.741 & 1107 & 35 & \bfseries 58 \\
LLaMA 4 Maverick 17B
  & \bfseries 0.859 & 0.834 & \bfseries 0.758 & 931 & \bfseries 200 & 69 \\
GPT-4o
  & 0.857 & \bfseries 0.838 & 0.774 & 945 & 197 & \bfseries 58 \\
\bottomrule
\end{tabular}
\end{adjustbox}
\end{table}

\section{Conclusion}
Standard GNN evaluations often overlook performance degradation from distribution drift and employ unrealistic adversarial attacks. Our analysis using a unified dataset confirms GNN susceptibility to drift, with further degradation under simulated realistic attacks. This work demonstrates that integrating \textit{individual} LLM as expert analyzers enhances GNN resilience against attacks like node injection, showcasing the potential of hybrid GNN-LLM systems.

Future work will focus on developing an LLM agentic pipeline. This pipeline aims to leverage different models strategically: for instance, employing GPT-4o for complex, high-level graph analysis to generate context that guides more cost-efficient models, such as Claude 3.5 Haiku, for scaled processing. The goal is to create a cost-effective workflow that maintains strong detection performance against network intrusions and adversarial attacks. Furthermore, we plan to validate these findings against diverse, realistic attack scenarios within a physical IoT testbed, utilizing real-world tools (e.g., Kali Linux) and incorporating physical device manipulations. Optimizing the overall pipeline efficiency for potential real-time NIDS deployment remains a crucial final objective.

\bibliographystyle{IEEEtran}
\bibliography{refer}

\end{document}